\documentclass{ws-procs9x6}
\newcommand{\be}{\begin{equation}}
\newcommand{\ee}{\end{equation}}

\newcommand{\ba}{\begin{eqnarray}}
\newcommand{\ea}{\end{eqnarray}}

\def\citebk#1{[\hspace{0.3mm}\raisebox{-1.85mm}[0mm][0mm]
  {\Large\cite{#1}}\hspace{-0.1mm}]}

\begin{document}

\title{Impact of CFL Quark Matter on the Cooling of Compact Stars}

\author{I.~A. SHOVKOVY$^{\lowercase{a,}}$\footnote{\uppercase{W}ork 
supported by \uppercase{G}esellschaft f\"{u}r
\uppercase{S}chwerionenforschung \uppercase{(GSI)} and by 
\uppercase{B}undesministerium f\"{u}r \uppercase{B}ildung 
und \uppercase{F}orschung \uppercase{(BMBF)}. \uppercase{O}n leave
of absence from \uppercase{B}ogolyubov \uppercase{I}nstitute for
\uppercase{T}heoretical \uppercase{P}hysics,
03143, \uppercase{K}iev, \uppercase{U}kraine.}
~\lowercase{and} 
P.~J. ELLIS$^{\lowercase{b,}}$\footnote{\uppercase{W}ork supported by 
the \uppercase{U.S. D}epartment of \uppercase{E}nergy \uppercase{G}rant 
\uppercase{N}o.~\uppercase{DE-FG02-87ER40328}.}}

\address{
$^{a}$Institut f\"ur Theoretische Physik, 
Johann Wolfgang Goethe--Universit\"at, \\
60054 Frankfurt/Main, Germany\\[1mm]
$^{b}$School of Physics and Astronomy, 
University of Minnesota, \\
Minneapolis, Minnesota 55455, USA}

\maketitle

\abstracts{The cooling mechanism of compact stars with quark cores in the
color-flavor locked phase is discussed. It is argued that the high thermal
conductivity of the quark core plays a key role in the stellar cooling. It
implies that the cooling time of compact stars with color-flavor locked
quark cores is similar to that of ordinary neutron stars, unless the star
is almost completely made of color-flavor locked quark matter.}
 
The observational study of compact stars is of prime importance because it
could potentially reveal the existence of new phases of matter at high
densities and low temperatures. Given the expected central densities it
was suggested long ago that some compact stars may be, at least partially,
made of quark matter.\cite{quark-stars} Recent observations on the
cooling\cite{quark1} of one neutron star and the radius\cite{quark2} of
another led the authors to suggest that exotic components, possibly
quarks, were required. These suggestions have been
disputed.\cite{no-quark1,no-quark2} However the current situation, where
there is no unambiguous evidence that free quarks play a role in compact
stars, may well change as further observations are made.

If the central densities of compact stars are indeed sufficient to support
quark matter, it is likely that it will be found in the color-flavor
locked (CFL) phase.\cite{ARW} (It has recently been suggested\cite{HS}
that two flavor color superconductivity may also play a role, but
we do not consider this possibility further here.) In the CFL phase the
three lightest flavors of quarks participate in a color condensate on an
approximately equal footing.  From a theoretical viewpoint there already
exists a rather detailed understanding of the basic properties of CFL
quark matter.\cite{ARW,CasGat,SonSt,ShoWij,Risch,other,rev} Here we
discuss the thermal conductivity of the CFL phase. Since the neutrino and
photon emission rates for the CFL phase are very small,\cite{JPS} the
thermal conductivity is expected to play the dominant role in the cooling
of stars with quark cores.\cite{cool-star,opaque}

The breaking of chiral symmetry in the CFL ground state leads to the
appearance of an octet of pseudo-Nambu-Goldstone (NG)  
bosons.\cite{ARW,CasGat,SonSt,other} In addition an extra NG boson $\phi$
and a pseudo-NG boson $\eta^{\prime}$ appear in the low energy spectrum as
a result of the breaking of global baryon number symmetry and approximate
$U(1)_{A}$ symmetry, respectively. The general structure of the low energy
action in the CFL phase can be established by symmetry arguments
alone.\cite{CasGat} However, the values of the parameters in such an
action can be rigorously derived only at asymptotically large baryon
densities.\cite{SonSt} Thus, in the most interesting case of intermediate
densities existing in the cores of compact stars the details of the action
are not well known. For some studies, however, it suffices to know that
there are 9 massive pseudo-NG bosons and one massless NG boson $\phi$ in
the low energy spectrum. At low temperatures, the NG bosons along with
photons should dominate the kinetic properties of dense quark
matter.\cite{cool-star} Some kinetic properties are also affected by the
presence of thermally excited electron-positron pairs.\cite{opaque}

At zero temperature, the electrical neutrality of the CFL phase is
enforced without the use of electrons.\cite{neutral} This may suggest that
the chemical potential of the electric charge, $\mu_e$, is zero also at
finite temperature. Then the densities of thermally excited electrons and
positrons would be equal. In fact Lorentz invariance is broken in the CFL
phase and the positively and negatively charged pions, as well as kaons,
differ in mass.\cite{bs} As a result, at finite temperature the densities
of the positively and negatively charged species are not exactly the same
and this must be balanced by differing electron and positron densities in
order to maintain charge neutrality. Thus $\mu_e$ is not exactly zero. At
temperatures below $5$ MeV, however, $\mu_e$ drops rapidly to zero and it
is completely negligible at temperatures of $1$ MeV or less which are our
principal interest here.\cite{rst} It is therefore sufficient to set
$\mu_e=0$ in assessing the impact of electron-positron pairs on the
physical properties of CFL quark matter.

The first issue to address is the photon mean free path. Since photons
scatter quite efficiently from charged leptons, even small numbers of
electrons and positrons could substantially reduce the transparency of CFL
quark matter. The photon mean free path can be rather well approximated by
the simple expression:
\be
\ell_{\gamma} \simeq \frac{1}{2 n_{e}(T) \sigma_{\rm T}}\;,
\ee
where $n_{e}(T)$ is the equilibrium density of electrons at temperature
$T$, the factor of 2 takes into account the equal density of positrons 
and
\be
\sigma_{\rm T} = \frac{8\pi}{3} \frac{\alpha^{2}}{m_{e}^{2}}
\approx 66.54 \mbox{~fm}^{2}
\ee
is the well-known expression for the Thomson cross section in terms of 
the fine structure constant $\alpha$ and the electron mass, $m_e$. This
expression is the limiting case of the more complicated Compton cross
section for low photon energies. Since this limit works rather well for
$\omega_{\gamma} \ll m_{e}$ and $m_{e} \simeq 0.5$ MeV, this is sufficient
for our purposes.

The average equilibrium electron or positron density in a 
finite temperature neutral plasma is:\cite{thermo-asympt}
\be
n_e = \frac{m_{e}^2 T}{\pi^{2}} \sum_{k=1}^{\infty}
\frac{(-1)^{k+1}}{k} K_{2}\!\left(\frac{m_{e}k}{T}\right),
\label{n_e}
\ee
where $K_2$ is a modified Bessel function. By making use of this result we
find that in the CFL phase the photon mean free path $\ell_{\gamma}
\lesssim 220$ m for temperatures $T \gtrsim 25$ keV. Since the radius of
the CFL core, $R_0$, is of order 1 km the photon mean free path is short
for temperatures above $25$ keV, meaning that the CFL quark core of a
compact star is opaque to light. Conversely, transparency sets in when the
mean free path exceeds $1$ km which occurs for temperatures below $23.4$
keV. In this case $\ell_\gamma\sim R_0$ because the photons are reflected
from the boundary with the nuclear matter layer due to the electron plasma
there.\cite{cool-star}

It is clear that at sufficiently low temperature (when the photon mean
free path exceeds $R_0$) massless photons and NG bosons, $\phi$, give
roughly equal contributions to the thermal conductivity of CFL matter. At
higher temperatures (i.e., $T \gtrsim 25$ keV), the photon contribution to
the thermal conductivity becomes negligible and only the contribution of
massless NG bosons needs to be considered. The corresponding mean free
path of the NG bosons is determined by their self-interactions, as well as
their possible dissociation into constituent quarks. Estimates of these
mean free paths are\cite{cool-star}
\ba
\ell_{\phi\phi\to\phi\phi} \sim \frac{\mu^8}{T^{9}}
\approx 8\times 10^5\frac{\mu_{500}^8}{T_{\rm MeV}^{9}}\mbox{ km},\\
\ell_{\phi\to q q} \sim \frac{v}{T} \exp\left(\sqrt{\frac{3}{2}}
\frac{\Delta}{T}\right)\;,         
\ea
respectively. Here we used the following notation: the quark chemical
potential $\mu_{500} \equiv\mu/(500\mbox{ MeV})$, $T_{\rm MeV} \equiv
T/(1\mbox{ MeV})$, the superconducting gap is denoted by $\Delta$
($\sim50$ MeV) and $v=1/\sqrt{3}$ is the velocity of the bosons $\phi$.
For temperatures $T\lesssim 1$ MeV, both $\ell_{\phi\phi\to\phi\phi}$ and
$\ell_{\phi\to qq}$ are much larger than the typical quark core radius.
Thus, it is only the geometry of the core, where the NG bosons $\phi$
exist, that limits the mean free path, giving $\ell \sim R_{0}$.

We note that the contribution of massive NG bosons to the thermal
conductivity is exponentially suppressed by a factor $\exp(-m/T)$ and the
contribution of quarks will likewise be suppressed due to the gap
$\Delta$. Then using the fact that $\ell \sim R_{0}$, the following
estimates are derived for the thermal conductivity of CFL matter inside 
a compact star:\cite{cool-star,opaque}
\be
\kappa_{\rm CFL} = \kappa_{\phi} + \kappa_{\gamma} 
\simeq \frac{2\pi^{2}}{9} T^{3} R_{0}\;, \quad \mbox{for} \quad 
T \lesssim 25\mbox{~keV}\;, 
\ee
\be
\kappa_{\rm CFL} = \kappa_{\phi} 
\simeq \frac{2\pi^{2}}{15} T^{3} R_{0}\;,
\quad \mbox{for} \quad T \gtrsim 25\mbox{~keV}\;.
\label{kappa-CFL}
\ee
Numerically, this gives
\be
\kappa_{\rm CFL} \simeq 1.2 \times 10^{32} T_{\rm MeV}^{3} R_{0,{\rm km}}
\mbox{~erg~cm}^{-1} \mbox{~sec}^{-1} \mbox{~K}^{-1}\;,\ \ \mbox{for}\ \ 
T \lesssim 25\mbox{~keV}\;,
\ee
\be
\kappa_{\rm CFL} \simeq 7.2 \times 10^{31} T_{\rm MeV}^{3} R_{0,{\rm km}}
\mbox{~erg~cm}^{-1} \mbox{~sec}^{-1} \mbox{~K}^{-1}\;,\ \ \mbox{for}\ \ 
T \gtrsim 25\mbox{~keV}\;,
\label{kappa-CFL-num}
\ee
where $R_{0,km}$ is the quark core radius measured in kilometers. This is
many orders of magnitude larger than the thermal conductivity of regular
nuclear matter in a neutron star.\cite{van} It is sufficient to wash away
a temperature gradient of 1 MeV across a 1 km core in a fraction of a
second.

In determining the cooling time an important role is obviously played by
the magnitude of the thermal energy which needs to be removed. There are
contributions to the total thermal energy from both the quark and the
nuclear parts of the star. The dominant amount of thermal energy in the
CFL core is stored in photons and massless NG bosons which
total,\cite{cool-star}
\be
E_{CFL}(T) 
=\frac{6(1+2v^3)T}{5} \left(\frac{\pi TR_{0}}{3v}\right)^{3},
\label{E_CFL}
\ee
where, for simplicity, the Newtonian approximation is used. 
Numerically this yields
\be
E_{CFL}(T) \simeq 2.1 \times 10^{42} R_{0,km}^{3} T_{\rm MeV}^{4}
\mbox{ erg}\;.\label{then}
\ee
The only other potentially relevant contribution to the thermal 
energy comes from electron-positron pairs. However, as shown in 
Ref.~\citebk{opaque}, this contribution is rather small in
comparison to Eq.~(\ref{then}) for the temperatures of interest here.
As regards the outer nuclear layer the thermal energy is provided 
mostly by degenerate neutrons. A numerical estimate is\cite{Shapiro}
\be
E_{NM}(T) \simeq 8.1 \times 10^{49} \frac{M-M_{0}}{M_{\odot}}
\left(\frac{\rho}{\rho_0}\right)^{\!-2/3} T^{2}_{\rm MeV}
\mbox{ erg}\;,
\ee
where $M$ is the mass of the star, $M_{0}$ is the mass of the quark core,
$M_{\odot}$ is the mass of the Sun and $\rho/\rho_0$ is the ratio of
the average nuclear matter density to equilibrium density. It is crucial 
to note that the thermal energy of the quark core is negligible in 
comparison to that of the nuclear layer. Moreover, as the star cools the 
ratio $E_{CFL}/E_{NM}$ will further decrease.

The second important component that determines stellar cooling is the
neutrino and/or photon luminosity which describes the rate of energy loss.
Typically, the neutrino luminosity dominates the cooling of young stars
when the temperatures are still higher than about $10$ keV and after that
the photon diffusion mechanism starts to dominate. It was argued\cite{JPS}
that neutrino emission from the CFL quark phase is strongly suppressed at
low temperatures, $T\lesssim 1$ MeV. The neighboring nuclear layer, on the
other hand, emits neutrinos quite efficiently. The nuclear layer should be
able to emit not only its own thermal energy, but also that of the quark
core which constantly arrives by the very efficient heat conduction
process. The analysis of this cooling mechanism, however, is greatly
simplified by the fact that the thermal energy of the quark core is
negligible compared to the energy stored in the nuclear matter. By making
use of the natural assumption that local neutrino emissivities from the
nuclear matter are not affected by the presence of the quark core, we
conclude that the cooling time of a star with a quark core by neutrinos is
essentially the same as for an ordinary neutron star provided that the
nuclear layer is not extremely thin.

After the neutrino cooling mechanism exhausts itself with the aging of a
star and surface cooling by photons starts to dominate (say, after about
$10^{5}$ years), the cooling rate of a star with a CFL core should become
faster than the corresponding rate for ordinary neutron stars. This
follows because only the nuclear matter layer of the hybrid star has a
sizable amount of the thermal energy that needs to be emitted through
the surface and this energy is only a fraction of the thermal energy in an
ordinary neutron star.

A completely different situation could arise if a compact star were made
of pure CFL quark matter. One could imagine that such a star may or may
not have a thin nuclear crust on the surface.\cite{quark-stars} If such
stars exist, their cooling would be unusual. The important fact about bare
CFL quark stars is that their neutrino and photon emissivities are
low\cite{JPS} and initially they have relatively little thermal energy.
Most of this energy is stored in $\phi$ bosons which are trapped. Photons
would also be present in the interior until the temperature dropped below
25 keV when they would all escape.

The cooling of a star such as this would largely depend on how well heat
conducted from the interior could be emitted from the surface in the form
of photons. If the photon emissivity were low, the star would shine dimly
for a long time. On the other hand if the emissivity were high, the
thermal energy conducted to the surface would be emitted rather rapidly
leaving the star cold and unseen. Apart, possibly, from a burst of photons
when the star became transparent, the prospects for detection of a star of
this type by anything other than its gravitational pull would seem slight.

In conclusion, the thermal conductivity of CFL matter in the core of a
compact star is large due to the presence of massless NG bosons, $\phi$.
Photons can increase the conductivity by a factor of 15/9 provided the
temperature is below about $25$ keV and the CFL quark core of the star is
surrounded by a reflective nuclear layer. Above this temperature, however,
the photon contribution is suppressed due to the presence of a thermally
excited electron-positron plasma inside the neutral CFL core. In either
case the conductivity of CFL matter is sufficiently large that the core
remains nearly isothermal. The thermal energy is efficiently conducted
away to the outer nuclear layer where it can be emitted in the form of
neutrinos and/or photons.\cite{cool-star}

\section*{Acknowledgments}

I.A.S. would like to thank the organizers for the invitation to attend the
Workshop. He is also grateful to Koichi Yamawaki and Masayasu Harada for
their kind hospitality.


\begin{thebibliography}{0}

\bibitem{quark-stars} E.~Witten,
{\it Phys. Rev.} {\bf D30}, 272 (1984);
C.~Alcock, E.~Farhi and A.~Olinto,
{\it Astrophys. J.} {\bf 310}, 261 (1986).

\bibitem{quark1} P.~Slane, D.J.~Helfand and S.S.~Murray,
{\it Astrophys. J.} {\bf 571}, L45 (2002).

\bibitem{quark2} J.J.~Drake, et al.,
{\it Astrophys. J.} {\bf 572}, 996 (2002).

\bibitem{no-quark1}
D.G.~Yakovlev, A.D.~Kaminker, P.~Haensel and O.Y.~Gnedin,
{\it Astron. \& Astrophys.} {\bf 389}, L24 (2002).

\bibitem{no-quark2} F.M.~Walter and J.M.~Lattimer,
{\it Astrophys. J.} {\bf 576}, L145 (2002).

\bibitem{ARW} M.~Alford, K.~Rajagopal and F.~Wilczek,
{\it Nucl. Phys.} {\bf B537}, 443 (1999).

\bibitem{HS} I.~Shovkovy and M.~Huang,
hep-ph/0302142;
I.~Shovkovy, M.~Hanauske and M.~Huang,
hep-ph/0303027.

\bibitem{CasGat} R.~Casalbuoni and R.~Gatto,
{\it Phys. Lett.} {\bf B464}, 111 (1999).

\bibitem{SonSt} D.T.~Son and M.A.~Stephanov,
{\it Phys. Rev.} {\bf D61}, 074012 (2000);
erratum {\em ibid.} {\bf D62}, 059902 (2000).

\bibitem{ShoWij} I.A.~Shovkovy and L.C.R.~Wijewardhana,
{\it Phys. Lett.} {\bf B470}, 189 (1999);
T.~Schafer,
{\it Nucl. Phys.} {\bf B575}, 269 (2000).

\bibitem{Risch} D.H.~Rischke,
{\it Phys. Rev.} {\bf D62}, 054017 (2000);
C.~Manuel and M.H.G.~Tytgat,
{\it Phys. Lett.} {\bf B501}, 200 (2001).


\bibitem{other} V.A.~Miransky, I.A.~Shovkovy and L.C.R.~Wijewardhana,
{\it Phys. Rev.} {\bf D63}, 056005 (2001);
V.~P.~Gusynin and I.~A.~Shovkovy,
{\it Nucl. Phys.} {\bf A700}, 577 (2002).

\bibitem{rev} K.~Rajagopal and F.~Wilczek,
{\it At the Frontier of Particle Physics: Handbook of QCD},
edited by M. Shifman (World Scientific, Singapore, 2001) Vol. 3, p.2061;
D.K.~Hong,
{\it Acta Phys. Polon.} {\bf B32}, 1253 (2001);
M.G.~Alford,
{\it Ann. Rev. Nucl. Part. Sci.} {\bf 51}, 131 (2001).

\bibitem{JPS} P.~Jaikumar, M.~Prakash and T.~Sch\"{a}fer,
{\it Phys. Rev.}  {\bf D66}, 063003 (2002).

\bibitem{cool-star} I.A.~Shovkovy and P.J.~Ellis,
{\it Phys. Rev.} {\bf C66}, 015802 (2002);
astro-ph/0207346.

\bibitem{opaque}  I.A.~Shovkovy and P.J.~Ellis,
{\it Phys. Rev.} {\bf C67}, (2003) in production, hep-ph/0211049.

\bibitem{neutral} K.~Rajagopal and F.~Wilczek,
{\it Phys. Rev. Lett.} {\bf 86}, 3492 (2001);
A.W.~Steiner, S.~Reddy and M.~Prakash,
{\it Phys. Rev.} {\bf D66}, 094007 (2002).

\bibitem{bs} P.F.~Bedaque and T.~Sch\"afer,
Nucl.\ Phys. {\bf A697}, 802 (2002).

\bibitem{rst} S.~Reddy, M.~Sadzikowski and M.~Tachibana,
Nucl. Phys. {\bf A714}, 337 (2003).

\bibitem{thermo-asympt} S.M.~Johns, P.J.~Ellis and J.M.~Lattimer,
{\it Astrophys. J.} {\bf 473}, 1020 (1996).

\bibitem{van} J.M. Lattimer, K.A. Van Riper, M. Prakash and 
M. Prakash, Astrophys. J.  {\bf425}, 802 (1994).

\bibitem{Shapiro} S.~L.~Shapiro and S.~A.~Teukolsky, {\sl Black
holes, white dwarfs, and neutron stars: the physics of compact
objects}, (John Wiley \& Sons, New York, 1983).

\end{thebibliography}
\end{document}